\documentclass[aps,floatfix,twocolumn]{revtex4-1}
\usepackage[pdftex]{graphicx}
\usepackage{amsmath}
\usepackage{hyperref}
\usepackage{latexsym}
\usepackage{color}
\usepackage{amssymb}
\usepackage{amsmath}
\usepackage [latin1]{inputenc}
\usepackage{bm}
\usepackage{graphicx}
\usepackage{siunitx}
\usepackage{amsmath,amssymb}
\usepackage{braket}
\usepackage[final]{microtype}

\begin{document}

\title{Ultracold coherent control of molecular collisions at a F{\"o}rster resonance}
\author{Thibault Delarue}
\affiliation{Universit\'{e} Paris-Saclay, CNRS, Laboratoire Aim\'{e} Cotton, 91405 Orsay, France}
\author{Goulven Qu{\'e}m{\'e}ner}
\affiliation{Universit\'{e} Paris-Saclay, CNRS, Laboratoire Aim\'{e} Cotton, 91405 Orsay, France}

\begin{abstract}
We show that the precise preparation of a quantum superposition 
between three rotational states of an ultracold dipolar molecule generates
controllable interferences in their
two-body scattering dynamics and collisional rate coefficients,
at an electric field that produces a F{\"o}rster resonance.
This proposal represents a feasible protocol to achieve coherent control 
on ultracold molecular collisions in current experiments. 
It sets the basis for future studies in which one can think 
to control the amount of each produced pairs, including trapped entangled pairs of reactants, 
individual pairs of products in a chemical reaction, 
and measuring each of their scattering phase-shifts that could 
envision ``complete chemical experiments" at ultracold temperatures.
\end{abstract}

\maketitle

The advent of ultracold, controlled dipolar molecules
has opened many exciting perspectives for the field of ultracold matter.
Their extremely controllable properties have inspired many theoretical proposals for promising quantum applications, such as quantum simulation and quantum information processes, quantum-controlled chemistry and test of fundamental laws \cite{Bohn_S_357_1002_2017,DeMille_S_357_990_2017}.
The molecules can be well prepared in individual quantum states \cite{Ni_S_322_231_2008},
their long-range interactions can be controlled \cite{Quemener_CR_112_4949_2012}, 
they can be long-lived and protected from their environment 
\cite{Micheli_PRL_105_073202_2010,Quemener_PRA_81_060701_2010,
DeMiranda_NP_7_502_2011,Chotia_PRL_108_080405_2012,Frisch_PRL_115_203201_2015,
Wang_NJP_17_035015_2015,Karman_PRL_121_163401_2018,Lassabliere_PRL_121_163402_2018,
Xie_PRL_125_153202_2020,Karam_PRR_5_033074_2023,
Matsuda_S_370_1324_2020,Li_NP_17_1144_2021,Anderegg_S_373_779_2021,
Schindewolf_N_607_677_2022,Bigagli_NP_19_1579_2023,Lin_PRX_13_031032_2023}, 
enabling the formation of quantum degenerate gases  \cite{DeMarco_S_363_853_2019,Valtolina_N_588_239_2020,Schindewolf_N_607_677_2022,Bigagli_arXiv_2312_10965_2023}, 
they can be manipulated in optical lattices
\cite{Yan_N_501_521_2013,Christakis_N_614_64_2023}
or in optical tweezers \cite{Liu_S_360_900_2018,Anderegg_S_365_1156_2019,Ruttley_PRL_130_223401_2023},
they can be used to explore many-body effects \cite{Micheli_NP_2_341_2006,Buchler_PRL_98_060404_2007,Gorshkov_PRL_107_115301_2011,
Baranov_CR_112_5012_2012,Yao_NP_14_405_2018,Schmidt_PRR_4_013235_2022},
they can be electro-associated to form long-range tetramer molecules
\cite{Quemener_PRL_131_043402_2023,Chen_arXiv_2306_00962_2023},
and they can be entangled
\cite{Holland_arXiv_2210_06309_2022,Bao_arXiv_2211_09780_2022}. 
Ultracold molecules can also be used to probe chemical reactions 
with an unprecedented control at the quantum level, 
as was done with the chemical reaction KRb + KRb $\to$ K$_2$ + Rb$_2$ at ultracold temperatures
\cite{Hu_S_366_1111_2019,Liu_N_593_379_2021}, 
including the control of the rotational parity 
of the products \cite{Hu_NC_13_435_2021,Quemener_PRA_104_052817_2021} 
and the creation of entangled product pairs \cite{Liu_arXiv_2310_07620_2023}.

In this study, we propose to apply the ideas of coherent control
\cite{Shapiro_PRL_77_2574_1996,Brumer_JCP_113_2053_2000,Shapiro_RPP_66_859_2003,Gong_JCP_118_2626_2003,
Devolder_PRL_126_153403_2021,Devolder_PRA_105_052808_2022,Devolder_PRR_5_L042025_2023,
Devolder_JPCL_14_2171_2023}
to current experiments of ultracold chemical reactions
\cite{Matsuda_S_370_1324_2020,Li_NP_17_1144_2021,Liu_N_593_379_2021}.
By using a microwave to prepare ultracold dipolar molecules in a quantum superposition of 
three stationary states (qutrit) and by using a static electric field to make collisional states degenerate, 
we predict that one can observe interferences in the rate coefficients of ultracold molecules.
This work provides a realistic and concrete experimental set-up for current experiments
to observe interferences and coherent control in ultracold collisions.

\begin{figure}[t]
\begin{center}
\includegraphics*[width=8.7cm, trim=0cm 1.2cm 2.5cm 0cm]{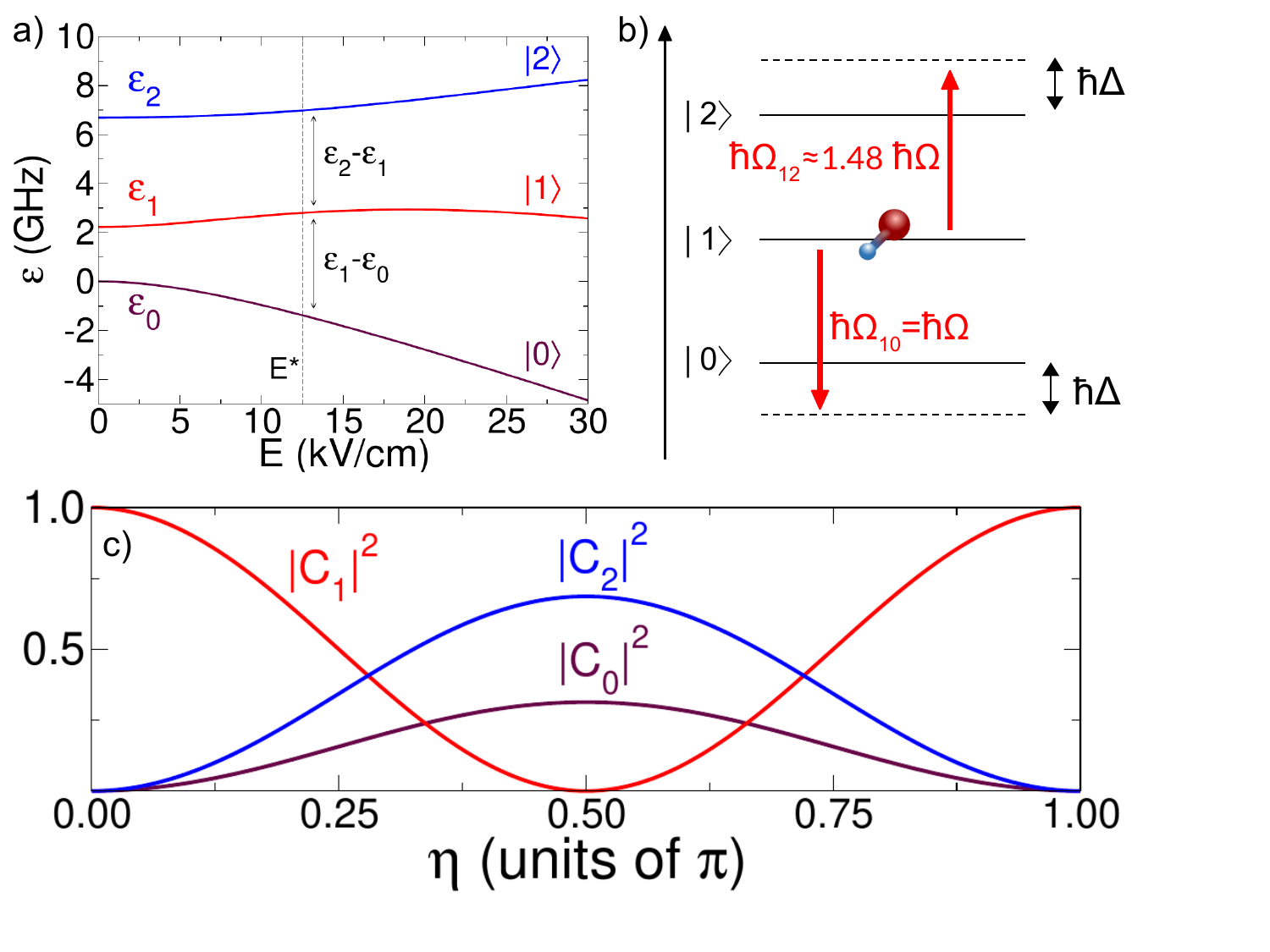}
\caption{ 
a) Energies $\varepsilon_0$, $\varepsilon_1$, $\varepsilon_2$ 
of the dressed states $\ket{0}$, $\ket{1}$, $\ket{2}$ as a function of the electric field $E$ (with all $m_j=0$). 
b) Sketch of the same energies at the F{\"o}rster resonance located at 
$E = E^* = 12.506$ kV/cm representing a ladder configuration in the presence of a microwave.
As the levels are equally spaced, the microwave creates a coupling between $\ket{0}$, $\ket{1}$ and $\ket{1}$, $\ket{2}$, with a detuning $\hbar \Delta$ and Rabi frequencies $\Omega_{01} \equiv \Omega$ and $\Omega_{12} \equiv \chi \, \Omega \approx 1.48 \, \Omega$. 
c) Evolution of the coefficient $|C_{0,1,2}|^2$ as a function of 
$\eta$ when $\Delta=0$. 
}
\label{FIG-1MOL}
\end{center}
\end{figure}

We first consider an ultracold dipolar gas of fermionic $^{40}\text{K}^{87}\text{Rb}$ molecules, taken as an example, in their ground electronic and vibrational state, and in their first excited rotational state. Their rotational states will be denoted by the kets $\ket{j,m_j}$, using usual notations for their quantum numbers $j$ and $m_j$. We omit the hyperfine structure and the nuclear spins of the atoms as they remain spectators at the magnetic fields considered in current experiments \cite{Matsuda_S_370_1324_2020,Li_NP_17_1144_2021}. 
In a static dc electric field $E$, the molecules are dressed into new states $\ket{\tilde{j},m_{j}}$
(also noted $\ket{j}$ for simplicity),
preserving the value of $m_j$. 
These are stationary states with well defined energies $\varepsilon_j$, 
as illustrated in Fig.~\ref{FIG-1MOL}a as a function of $E$.
Following the experiments in \cite{Matsuda_S_370_1324_2020,Li_NP_17_1144_2021}, 
we assume that all the molecules are prepared in a dressed state $\ket{\tilde{1},0}$, noted $\ket{1}$,
with energy $\varepsilon_1$ and we will assume a collision energy of $E_c / k_B = 500$~nK.
Two other states, namely $\ket{\tilde{0},0}$ (noted $\ket{0}$) with energy $\varepsilon_0$ 
and $\ket{\tilde{2},0}$ (noted $\ket{2}$) with energy $\varepsilon_2$, will be of interest
in the following.
We impose the electric field to be $E = E^* = 12.506$ kV/cm,
reachable in those experiments and for which 
$\varepsilon_1 - \varepsilon_0$ 
= $\varepsilon_2 - \varepsilon_1$.
This characterizes the position of 
a F{\"o}rster resonance \cite{Comparat_JOSAB_27_208_2010}.

Then, by applying a linearly polarized microwave of defined frequency and intensity
during a time $\tau$, 
we couple the states $\ket{0}$ and $\ket{2}$ to the state $\ket{1}$, 
and create Rabi oscillations between those three stationary states, consequently forming a qutrit.
This is illustrated in Fig.~\ref{FIG-1MOL}b.
The frequency of the microwave defines an equal detuning $\hbar \Delta$ 
as the energy spacings are the same, while its intensity defines an ac electric field $E_{ac}$, corresponding to two Rabi frequencies $\Omega_{10} = \tilde{d}^{1\leftrightarrow 0} \, E_{ac} / \hbar \equiv \Omega$ and $\Omega_{12} = \tilde{d}^{1\leftrightarrow 2} \, E_{ac} / \hbar \equiv \chi \, \Omega$, depending on the
electric dipole moments of the transitions. The quantities 
$\tilde{d}^{1\leftrightarrow 2} = 0.293$~D and $\tilde{d}^{1\leftrightarrow 0} = 0.198$~D 
are generalized induced dipole moments at $E^*$ \cite{Lassabliere_PRA_106_033311_2022}, 
the ratio of which gives $\chi \approx 1.48$.
This prepares a quantum superposition defined by the wavefunction at any time $t \ge \tau $
\begin{eqnarray}\label{qs1MOL}
 \ket{\Psi_{mol}}= \sum_{i=0}^2
 {c}_i \ket{i}  e^{\text{i} (\vec{k}_i.\vec{\rho} - \omega_i t)} 
 \equiv  \sum_{i=0}^2
 C_i \ket{i}  e^{\text{i} (\vec{k}_i.\vec{\rho} - \omega_i (t-\tau))} \nonumber \\
\end{eqnarray}
with ${c}_i = C_i \, e^{\text{i} \omega_i \tau}$, 
$\vec{k}_i$ and $\hbar \omega_i = \hbar^2 k_i^2 / 2m + \varepsilon_i$
are respectively the initial wavevector and energy of the individual molecule of mass $m$, described by a position vector $\vec{\rho}$, formed in states $i$. 
Due to negligible recoil energy and Doppler effect for a microwave transition \cite{SM},
the wavevector of the molecule when excited in state $\ket{2}$
or de-excited in state $\ket{0}$ is the same as the initial one in state $\ket{1}$,
namely $\vec{k}_{2} \simeq \vec{k}_{0} \simeq \vec{k}_{1}$.

Control of the interferences is reached through the 
control of this quantum superposition and the $C_{i}$ factors.
This is achieved by monitoring the parameters $\Delta$, $\Omega$ and $\tau$. 
At the F{\"o}rster resonance, 
the dynamics of the superposition is dictated by the resulting light-matter Hamiltonian.
To keep this work general, we will focus on the condition $\Delta=0$ for which the superposition factors 
are analytical. At time $t=\tau$, the microwave is turned off
and the $C_i$ factors are well defined \cite{SM} given by
$C_1 = \cos\eta$, $C_0 = - \text{i} \sin\Theta \sin\eta$, $C_2 = - \text{i} \cos\Theta \sin\eta$,
with $\eta = \sqrt{1+\chi^2} \, \Omega \, \tau/2$
and
$\tan\Theta = \Omega_{10}/\Omega_{12}=1/\chi \approx 0.68$.
The modulus square of these factors are represented in Fig.~\ref{FIG-1MOL}c.
The preparation time $\tau$ should be shorter than a typical collisional time
($\tau \ll \tau_{col} \simeq 1$~ms for KRb molecules at $E^*$).

\begin{figure}[t]
\begin{center}
\includegraphics*[width=8.7cm, trim=0cm 1.2cm 3cm 0cm]{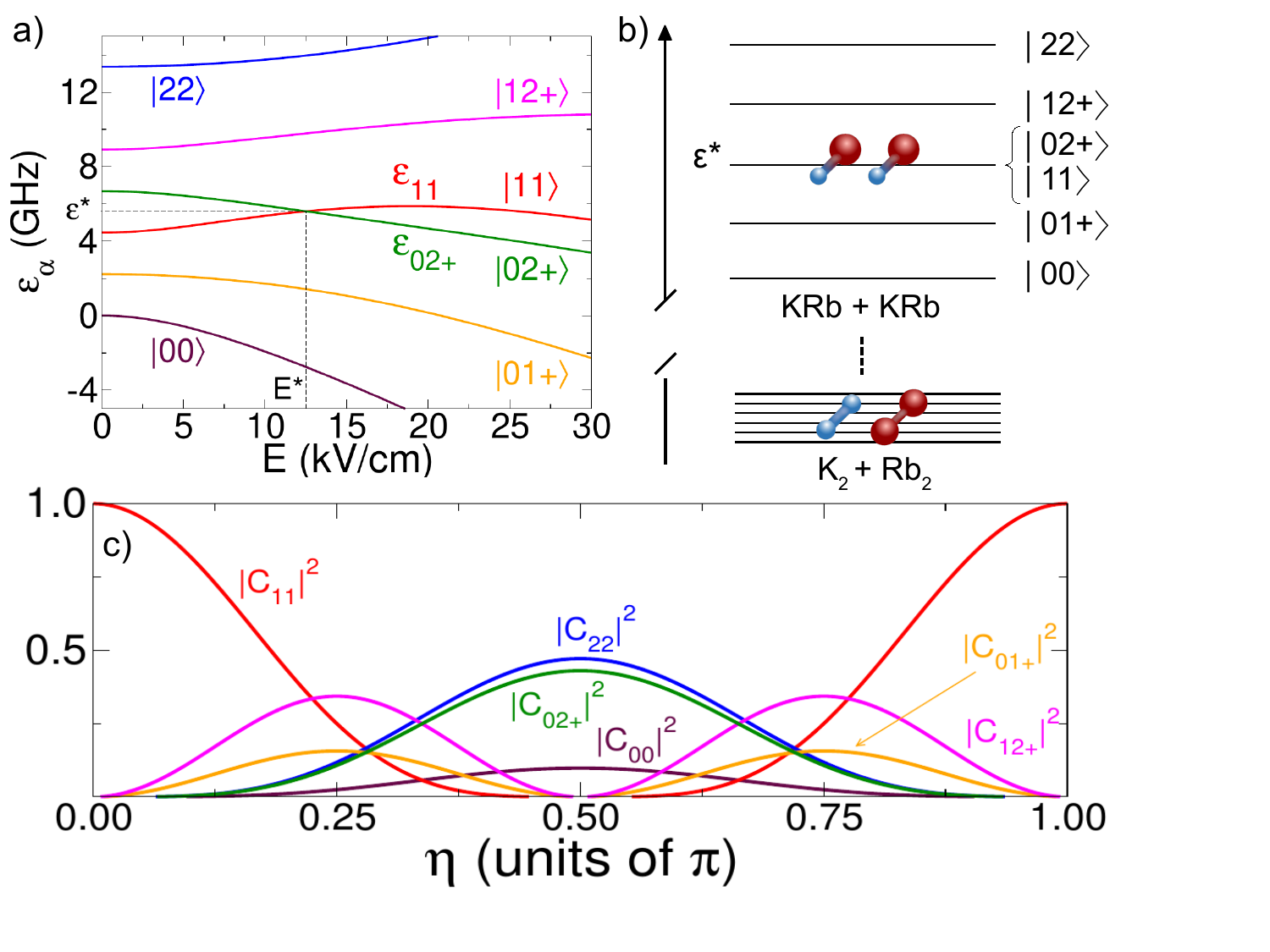}
\caption{ 
a) Energies $\varepsilon_\alpha$ with $\alpha = 00, 01+, 11, 02+, 12+, 22$ 
as a function of $E$ (with all $m_j=0$). At $E^*$, the energy $\varepsilon_{11} = \varepsilon_{02+} \equiv \varepsilon^*$. Not shown, energies for combined molecular states including non-zero values of $m_j$.
b) Sketch of the same energies at the F{\"o}rster resonance.
c) Evolution of the coefficient $|C_\alpha|^2$ as a function of $\eta$ when $\Delta=0$. 
}
\label{FIG-2MOL}
\end{center}
\end{figure}

Then after a time $t \ge \tau $, the molecules are free to collide.
The overall incident collisional wavefunction between two molecules
\cite{Brumer_JCP_113_2053_2000,Shapiro_RPP_66_859_2003}
prepared as in Eq.~\eqref{qs1MOL} 
is 
\begin{multline}\label{qs2MOLinc}
 \ket{\Psi_{col}^{inc}} = \ket{\Psi_{mol}} \otimes \ket{\Psi_{mol}} \\
 = \sum_{\alpha}^* C_{\alpha}  \ket{\alpha} e^{\text{i} \vec{k}_{{ \alpha }} . \vec{r} } 
   e^{\text{i} \vec{K}_{\alpha}.\vec{R}} e^{- \text{i} \omega_\alpha (t-\tau)} 
\end{multline}
where $\vec{k}_{{ \alpha }}$, $\vec{K}_{{ \alpha }}$ and $\hbar \omega_\alpha = \hbar^2 k_\alpha^2 / 2\mu + \hbar^2 K_\alpha^2 / 2 M + \varepsilon_\alpha$
are respectively the initial wavevector of the relative motion of reduced mass $\mu=m/2$ described by a position vector $\vec{r}$, the initial wavevector of the center-of-mass motion of total mass $M=2m$ described by a position vector $\vec{R}$, 
and the energy of the two molecules initialy prepared in 
one of the six possible combined molecular states 
$\ket{\alpha} \equiv \ket{00}, \ket{11}, \ket{22}, \ket{01+}, \ket{12+}, \ket{02+}$,
arising from the possible 
combinations of states $\ket{0}, \ket{1}, \ket{2}$ in Eq.~\eqref{qs1MOL}.
The internal energy $\varepsilon_\alpha$ of these states 
are plotted in Fig~\ref{FIG-2MOL}a as a function of $E$.
These combined molecular states are properly symmetrized under exchange of identical 
particles \cite{SM}.
The asterisk sign over the sum in Eq.~\eqref{qs2MOLinc} means that the states $\alpha$ are retricted
to the six ones mentioned above.
The $C_{\alpha}$ factors are given by 
$C_{11} = \cos^2\eta$, 
$C_{00} = -\sin^2\Theta\sin^2\eta$, 
$C_{22} = -\cos^2\Theta\sin^2\eta$,
$C_{01+} = -\text{i}\sqrt{2}\sin\Theta\sin\eta\cos\eta $, 
$C_{12+} = -\text{i}\sqrt{2}\cos\Theta\sin\eta\cos\eta  $,
$C_{02+} = -\sqrt{2}\sin\Theta\cos\Theta\sin^2\eta  $ \cite{SM},
the modulus square of which are represented in Fig~\ref{FIG-2MOL}c.

Eq.~\eqref{qs2MOLinc} is then a quantum superposition of six possible incident wavefunctions.
Each of them have a well defined energy $\hbar \omega_\alpha$
and produces a scattered wavefunction that we can compute \cite{Wang_NJP_17_035015_2015}
as the result of the collision. 
Note that the goal of the microwave preparation 
is not to keep the coherence of the individual molecules, 
as these scattered terms will definitively destroy the qutrits,
but rather to populate different molecular collisional states with 
the controlled $C_{\alpha}$ factors.
The asymptotic form of the general wavefunction of the system is then given by the quantum superposition
of different collisional wavefunctions
\begin{multline}\label{qs2MOL}
 \ket{\Psi_{col}} \underset{r \to \infty}{=} 
\sum_{\alpha}^*  \, \bigg[  C_{\alpha} \, \ket{\alpha} \, e^{\text{i} \vec{k}_{{ \alpha }} . \vec{r} } \\
 + {  \sum_{\alpha'} }  C_{\alpha} \,  f^{scat}_{{ \alpha \to \alpha'}}(\vec{k}_{ \alpha },\hat{r}) \, \frac{ e^{\text{i} k^{\alpha}_{\alpha'} r} }{r}  \ket{\alpha'} \bigg] 
  e^{\text{i} \vec{K}_{\alpha}.\vec{R}} \,  e^{- \text{i} \omega_\alpha (t-\tau)} 
\end{multline}
with the scattering amplitude given by
\begin{multline} 
f^{scat}_{\alpha \to \alpha'}(\vec{k}_\alpha,\hat{r}) = \frac{2 \pi}{\text{i} \, k_\alpha^{1/2} \, {k^{\alpha}_{\alpha'}}^{1/2}} \,  \sum_{l=0}^{\infty} \sum_{m_l=-l}^{l} \sum_{l'=0}^{\infty} \sum_{m_l'=-l'}^{l'}  \\
\text{i}^{l-l'} \, [Y_l^{m_l}(\hat{k}_\alpha)]^* \, Y_{l'}^{m_l'}(\hat{r}) \, T_{\alpha' \, l' \, m_l',\alpha \, l \, m_l}(k_\alpha)
\end{multline}
in term of elements of the transition matrix $T$ that we can compute,
employing an usual partial wave expansion over $l,m_l,l',m_l'$,
for a given total angular momentum projection quantum number $M$ 
\cite{Wang_NJP_17_035015_2015}.
The quantities $k^{\alpha}_{\alpha'}$ are found using
the conservation of energy after a collision
$\hbar^2 \, ({k^{\alpha}_{\alpha'}})^2 / 2 \mu + \varepsilon_{\alpha'}$ = 
$E_c + \varepsilon_\alpha$. 
Note that the states $\alpha'$ are all the possible combined molecular states to which 
the system can end up after a collision 
and are not restricted to the six ones prepared in Eq.~\eqref{qs2MOLinc}.
For example, molecules can end up in 
states with values of $m_j \ne 0$ such as $\ket{\tilde{1},\pm1}$, $\ket{\tilde{2},\pm1}$,
$\ket{\tilde{2},\pm2}$ and this is included in our study \cite{Wang_NJP_17_035015_2015}.

Because $\vec{k}_{2} \simeq \vec{k}_{0} \simeq \vec{k}_{1}$ as mentioned earlier in Eq.~\eqref{qs1MOL},
all the vectors $\vec{k}_{{\alpha}}$ are also equal \cite{SM} 
(we will note them $\vec{k}$) and the related collision energies
$\hbar^2 k_\alpha^2 / 2\mu$ are the same as the initial molecules in the gas, 
namely $E_c / k_B = 500$~nK. 
Similarly, all the vectors $\vec{K}_{{\alpha}}$ are equal  
(we will note them $\vec{K}$) so that the 
six possible centers-of-mass, where a collision can take place, move in the same way.
This is in fact a minimal requirement needed to expect for interferences 
between the six collisional waves in Eq.~\eqref{qs2MOL}.
However, if the angular frequencies $\omega_\alpha$
appearing in the time-dependent terms are different, 
the six collisional waves are not synchronous, that is they will not occur at the same time.
They will do so if and only if the combined molecular states $\ket{\alpha}$ 
have the same internal energies $\varepsilon_\alpha$
\cite{Brumer_JCP_113_2053_2000,Shapiro_RPP_66_859_2003}.
This is in general not an obvious requirement to fulfill.
But this can be done precisely for the states $\ket{11}$ and $\ket{02+}$
at $E=E^*$, by definition of the F{\"o}rster resonance,
where $\varepsilon_{11} = \varepsilon_{02+} = \varepsilon^*$,
as illustrated in Fig.~\ref{FIG-2MOL}b.
Then $\hbar \omega_{11} = \hbar \omega_{02+}$
(we note them $\hbar \omega$)
and $k^{11}_{\alpha'} = k^{02+}_{\alpha'}$ (we note them $k^{11/02+}_{\alpha'} $).
As the states $\ket{11}$ and $\ket{02+}$ collide
in the same center-of-mass and become synchronous, they can interfere. 
This is the key point of the paper.
We can factorize those two terms in Eq.~\eqref{qs2MOL} to get
the interference part of the collisional wavefunction
\begin{multline}\label{qs2MOL-INT}
\ket{\Psi_{col}^{int}} \underset{r \to \infty}{=} 
 \bigg\{ \bigg[ C_{11} \, \ket{11} + C_{02+} \, \ket{02+} \bigg]
  \, e^{\text{i} \vec{k}.\vec{r}} \\
+  \sum_{\alpha'}
 \bigg[ C_{11} \, f^{scat}_{11 \to \alpha'} + C_{02+} \, f^{scat}_{02+ \to \alpha'}\bigg] \, \ket{\alpha'}  \,  \frac{ e^{\text{i} k^{11/02+}_{\alpha'}  r} }{r} \bigg\} \\  
  e^{\text{i} \vec{K}.\vec{R}} \, e^{- \text{i} \omega (t-\tau)} .
\end{multline}
For the four remaining terms $\alpha = 00, 22, 01+, 12+$, also called ``satellite" terms \cite{Brumer_JCP_113_2053_2000,Shapiro_RPP_66_859_2003,Gong_JCP_118_2626_2003}, 
they do not interfere in Eq.~\eqref{qs2MOL} and will provide each of them an independent result. 
For a starting quantum superposition state, 
one can then define in Eq.~\eqref{qs2MOL-INT} new expressions of \cite{Brumer_JCP_113_2053_2000,Shapiro_RPP_66_859_2003,Gong_JCP_118_2626_2003,Devolder_PRL_126_153403_2021}
\begin{eqnarray}\label{FT-INT}
f^{qs}_{11/02+ \, \rightarrow\alpha'} &=& C_{11} f^{scat}_{11\rightarrow\alpha'}
+ C_{02+} f^{scat}_{02+\rightarrow\alpha'} \\
T^{qs}_{\alpha' l' m_l',11/02+ \, lm_l} &=& C_{11} T_{\alpha' l' m_l',11 \, lml} 
+ C_{02+} T_{\alpha' l' m_l',02+ \, lm_l} \nonumber 
\end{eqnarray}
where the notation $11/02+$ is used now to illustrate that $11$ and $02+$ 
are interfering and cannot be considered separately, and
\begin{align}\label{FT-SAT}
f^{qs}_{\alpha\rightarrow\alpha'} &= C_{\alpha} f^{scat}_{\alpha\rightarrow\alpha'}  &
T^{qs}_{\alpha' l'm_l',\alpha l m_l} &=
C_{\alpha} T_{\alpha'l'm_l',\alpha l ml}
\end{align}
for the satellite terms.
The rate coefficient ending in any combined molecular state $\alpha'$ is given by
\begin{eqnarray}\label{rate}
\beta^{qs}_{\alpha'}(E_c) = \sum_{\alpha}^*  \beta^{qs}_{\alpha\rightarrow\alpha'}(E_c)
\end{eqnarray}
where the sum runs over $\alpha = 11/02+$, $00$, $01+$, $12+$, $22$, with
\begin{eqnarray}\label{rate-s2s}
 \beta^{qs}_{\alpha\rightarrow\alpha'}(E_c) =  
 2 \, \frac{\pi \hbar}{\mu k}  \sum_l\sum_{m_l}\sum_{l'}\sum_{m_l'} \left|T^{qs}_{\alpha' l' m_l', \alpha l m_l}\right|^2. 
\end{eqnarray}
The rate coefficients depend parametrically on $\eta$ via the $C_{\alpha}$ coefficients 
in Eq.~\eqref{FT-INT} and Eq.~\eqref{FT-SAT}.
The factor of 2 takes into account that the initial molecules are indistinguishable.
The overall reactive rate coefficient is given by \cite{SM}
\begin{eqnarray}\label{rateloss}
 \beta^{qs}_{re}(E_c) = \sum_{\alpha}^* 2 \, \frac{\hbar \pi}{\mu k} \, |C_{\alpha}|^2 \,  \sum_l\sum_{m_l} P^{re}_{\alpha l m_l}   
\end{eqnarray}
where the sum runs over $\alpha =$ $11$, $02+$, $00$, $01+$, $12+$, $22$.

\begin{figure}[t]
\begin{center}
\includegraphics[width=8.7cm]{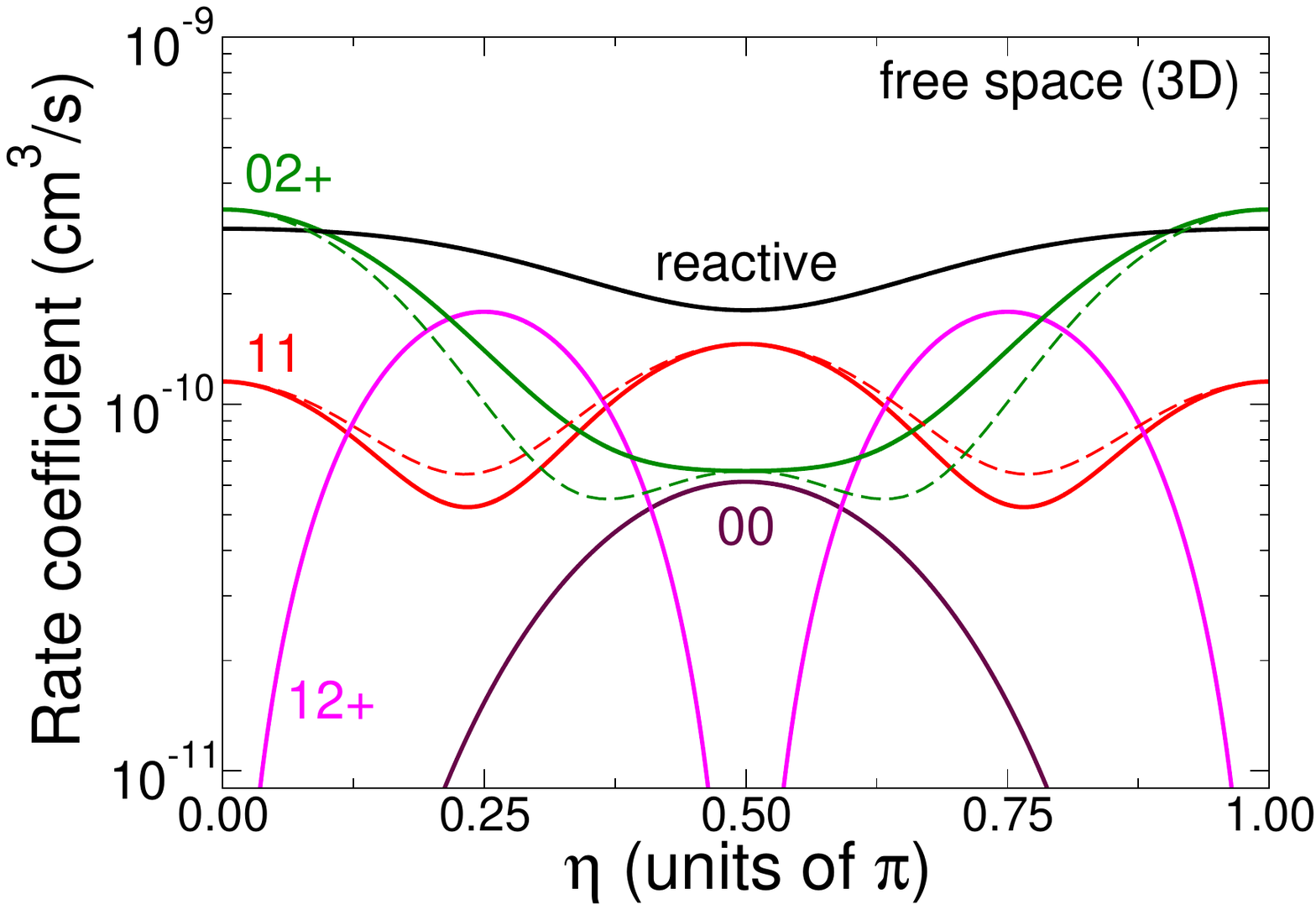} 
\caption{ 
Rate coefficients $\beta^{qs}_{\alpha'}$ of two qutrits at a F{\"o}rster resonnance as a function of the control parameter $\eta$, in a free-space 3D geometry at $E_c/k_B = 500$~nK. 
The solid red (resp. green) curve corresponds 
to a measurement in the final state $\alpha'=11$ (resp. $\alpha'=02+$). 
The dashed curves corresponds to the same curves but if there were no interferences.
The solid black curve corresponds to the overall reactive rate coefficient. The solid brown (resp. pink) curve corresponds to a measurement 
in the final state $\alpha'=00$ (resp. $\alpha'=12+$). 
}
\label{FIG-RATE-3D}
\end{center}
\end{figure}

\begin{figure}[t]
\begin{center}
\includegraphics[width=8.7cm]{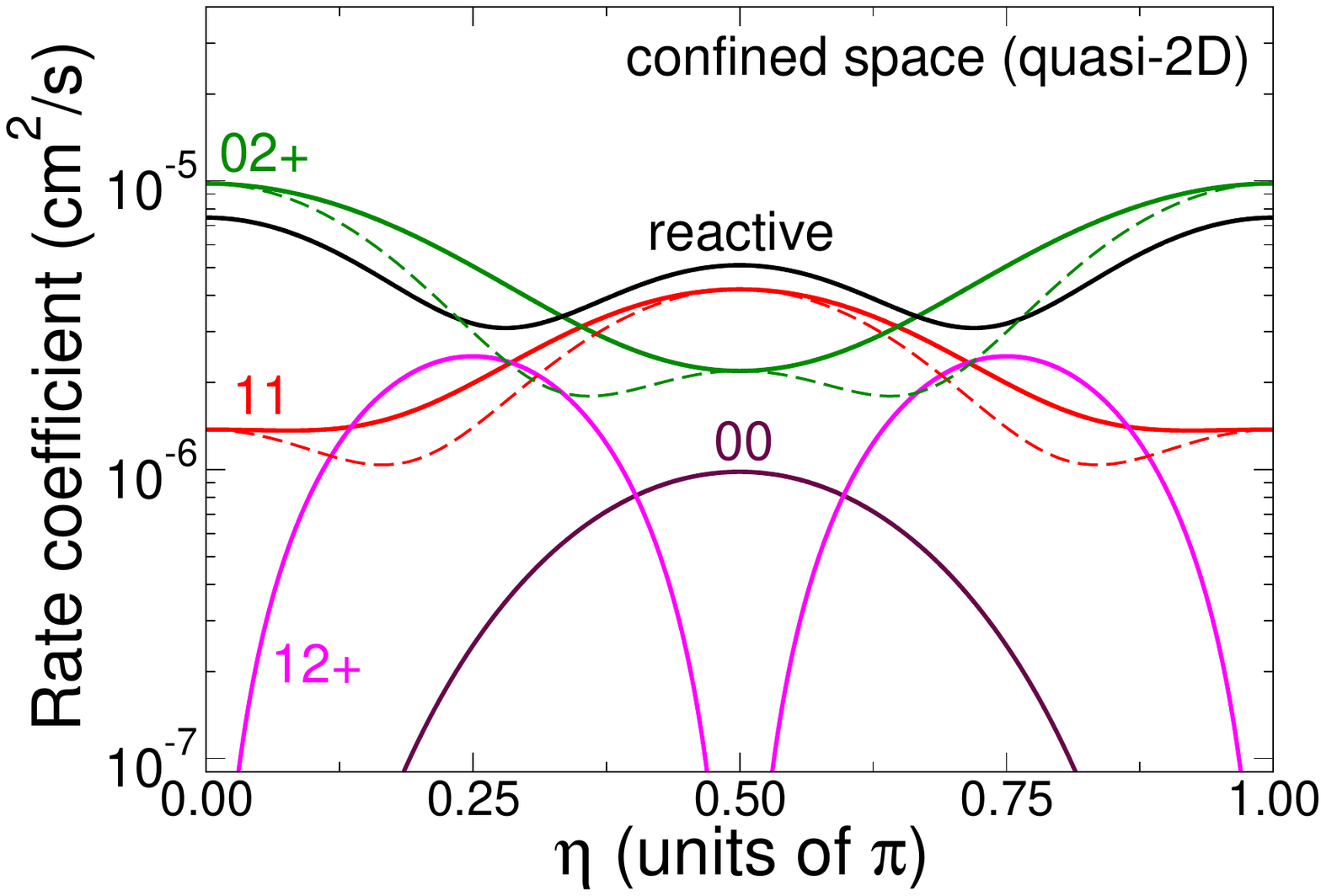} 
\caption{ 
Same as Fig.~\ref{FIG-RATE-3D} but in a confined quasi-2D geometry, using the same conditions
as in \cite{Matsuda_S_370_1324_2020}.
}
\label{FIG-RATE-2D}
\end{center}
\end{figure}

We computed all the $T$ and $P^{re}$ elements apearing in Eq.~\eqref{rate-s2s} and Eq.~\eqref{rateloss} 
at $E_c/k_B= 500$~nK, using the same basis sets as in \cite{Wang_NJP_17_035015_2015} that showed very good agreement with experimental observations 
in a free-space 3D geometry ($M=0,\pm1$, $l=1,3,5$, see conditions in \cite{Li_NP_17_1144_2021}) 
and in a confined quasi-2D geometry 
($M=\pm1$, $l=1,3,5$, see conditions in \cite{Matsuda_S_370_1324_2020}).
We plot the corresponding rate coefficients in Fig.~\eqref{FIG-RATE-3D} for the 
free-space case and in Fig.~\eqref{FIG-RATE-2D} for the confined case.
In all cases, the rate to a state $\alpha'$ in Eq.~\eqref{rate}
comes mainly from the contribution of the 
elastic term $\alpha=\alpha'$ in Eq.~\eqref{rate-s2s},
those for which the kinetic energy of the relative motion of the two molecules  
does not change \cite{Hutson_BookChapter_2009}.
It means that all the final reactants in a state $\alpha'$
have mainly the same final kinetic energies than their initial ones, 
so that they remain ultracold and still trapped.

First, we can see on both figures that the rates present variations as a function of $\eta$.
Though, for $\alpha'=00$ (solid brown curves) and $12+$ (solid pink curves), 
they do not correspond to interferences 
as can be seen from Eq.~\eqref{FT-SAT}.
The rates are just respectively proportional to $|C_{00}|^2$ and $|C_{12+}|^2$ 
in Eq.~\eqref{rate-s2s}.
They just exhibit the same $\eta$ behaviour as those preparation coefficients,
as can be seen in Fig.~\ref{FIG-2MOL}c.
The rates to $\alpha'=01+$, $22+$, as well as to the other combined molecular states
involving $m_j \ne 0$ are much smaller and do not appear in the figure. 
For $\alpha'=11$ (solid red curves) and $02+$ (solid green curves) for the free-space case, 
the variations correspond to 
destructive and constructive interferences respectively, when compared
to the same rates (dashed curves) 
without the crossed interference term 
taken into account in Eq.~\eqref{rate} and Eq.~\eqref{FT-INT}.
By controlling individually 
$C_{11}$ and $C_{02+}$ with $\eta$, one coherently controls 
the scattering amplitude and the $T$ matrix elements, 
hence the observables
$\beta^{qs}_{11}$ and $\beta^{qs}_{02+}$, with changes of magnitude up to a factor of five here. 
For the confined case, similar conclusions hold but now both states $\alpha'=11$ and $02+$ 
correspond to constructive interferences.

Secondly, the rich physics at this F{\"o}rster resonance 
is a remarkable source for ultracold entangled pairs
remaining in the trap.
As can be seen for $\eta=0$ 
(conditions already fulfilled in \cite{Matsuda_S_370_1324_2020,Li_NP_17_1144_2021}),
we predict a large production in $\alpha'=02+$, 
namely trapped pairs in the entangled states $\ket{02+} = \{\ket{02}+\ket{20}\}/\sqrt{2}$,
even larger than the production in the non-entangled trapped pairs $\alpha'=11$, 
or in the non-trapped reactive pairs. 
When $\eta$ increases from 0 to $\pi/2$, 
it turns out that $\beta^{qs}_{02+}$ decreases in a monotonic way.
Even though interferences are constructive,
the fact that the coefficients $C_{11}$ and $C_{02+}$ 
are purely real numbers for $\Delta = 0$ (without additional phases to monitor),
decreases somewhat the flexibility to control the amount of interferences in Eq.~\eqref{FT-INT}.
However, if we consider the case $\Delta \ne 0$,
the $C_{11}$ and $C_{02+}$ coefficients 
become arbitrary complex numbers (including two phases to monitor) that can be controlled 
in many ways by the three independent values $\Delta$, $\Omega$, and $\tau$.
This can be used to control the amount of interferences, for example
to increase even more the production of entangled states.
Coherent control could then become a tool to enhance the production of trapped entangled pairs
and this will be left for a future study.

Finally, the overall reactive rate coefficients, while showing variations with the control parameter $\eta$, 
do not exhibit signatures of interferences, as it results in the sum of many 
state-to-state contributions of the possible reactive product pairs, included
in a phenomenological way 
through $P^{re}_{\alpha l m_l}$ in Eq.~\eqref{rateloss} \cite{Wang_NJP_17_035015_2015,SM}.
This is due to the fact that we cannot theoretically predict the 
individual state-to-state reactive rates, because we cannot predict
the complex-valued $T$ matrix elements of each of the products pairs with our formalism.
Only the norms of those elements are known and observed \cite{Liu_N_593_379_2021}, 
seemingly consistent with a statistical model
\cite{Gonzalez-Martinez_PRA_90_052716_2014,Bonnet_JCP_152_084117_2020,Huang_JPCA_125_6198_2021}, 
but the individual phases remain still unknown.
However, an experimental set-up as in \cite{Liu_N_593_379_2021}
combined with the one in \cite{Matsuda_S_370_1324_2020,Li_NP_17_1144_2021}, 
would be able to directly measure 
those state-to-state interfering reactive probabilities as a function of $\eta$. 
By fitting the experimental results, one could imagine to get back to all 
the state-to-state complex-valued $T$ matrix elements (including the phases)
of the $\alpha=11$ and $\alpha=02+$ columns in Eq.~\eqref{FT-INT}.
This is reminiscient of a ``complete chemical experiment" \cite{Bohn_S_357_1002_2017}
at ultracold temperature, here from an interferometry-type experiment between two collisional waves of matter, 
with the prospect of measuring all the scattering phase-shifts of a reaction.

In conclusion, we showed that a quantum-controlled microwave preparation 
of a dipolar molecule at a F{\"o}rster resonance in an electric field,
set the conditions for observing interferences between collisional waves
and coherent control of their dynamics, in current ultracold molecular experiments.
Evidences of constructive or destructive interferences
are predicted to appear in the rate coefficients
of two dipolar molecules measured in their first excited rotational state ($\alpha'=11$)
or in their ground and second excited rotational state ($\alpha'=02+$),
for different cases of confinements. 
The overall reactive rate coefficients do not exhibit signatures of interferences,
but the state-to-state reactive rates are expected to do so.
Future experimental observations of such individual rates 
in the distribution of the product pairs 
could be done combining the set-ups of Ref.~\cite{Liu_N_593_379_2021}
and Refs.~\cite{Matsuda_S_370_1324_2020,Li_NP_17_1144_2021}. 
This would be a signature of coherent control of ultracold chemistry at the state-to-state level.
This study opens many interesting perspectives as
coherent control could be used to enhance entanglement production
of ultracold trapped pairs needed for many quantum physics applications,
and to measure each individual state-to-state scattering phase shifts of a reaction
envisioning ``complete chemical experiments" at ultracold temperatures.

This work was supported by a grant overseen by the French National Research Agency (ANR) as part of the France 2030 program  ``Comp{\'e}tences et M{\'e}tiers d'Avenir" within QuantEdu France project reference: ANR-22-CMA-0001 and by Quantum Saclay center.

\bibliography{../../../BIBLIOGRAPHY/bibliography.bib}

\onecolumngrid

\section*{Supplemental material}

\subsection{Recoil energy and Doppler effect for a microwave transition}

The energy and the momentum of a molecule are not conserved during a stimulated microwave process, the difference being given by the energy and momentum of the emitted/absorbed photon.
Spontaneous emission processes can be neglected here for microwave transitions.
Let's consider a general case with a molecule $A$
in state $\ket{1}$ with wavevector $\vec{k}_{A,1}$
undergoing a stimulated emission process $\ket{1}\rightarrow \ket{0}$ 
in state $\ket{0}$ getting a wavevector $\vec{k}_{A,0}$,
with a photon $a$ of energy 
$\hbar \omega_{a}$ and momentum $ \vec{k}_{a}$.
From conservation of energy, we have
\begin{equation}\label{energy01}
\frac{\hbar^2k^2_{A,1}}{2m}+\varepsilon_{1}=\frac{\hbar^2k^2_{A,0}}{2m}+\varepsilon_{0}+\hbar\omega_A
\end{equation}
where $\hbar\omega_A$ is the energy of the photon emitted by the molecule $A$.
From conservation of momentum, we have 
\begin{eqnarray}\label{momentum01}
\hbar \vec{k}_{A,1} = \hbar \vec{k}_{A,0} + \hbar \vec{k}_{a} .
\end{eqnarray}
If we choose a space fixed frame such as the laboratory frame, we have using Eq.~\eqref{momentum01} 
\begin{eqnarray}\label{dev1}
\frac{\hbar^2 \vec{k}_{A,0}^2}{2m}  &=& \frac{\hbar^2 \big(\vec{k}_{A,1} - \vec{k}_{a}\big)^2 }{2m} 
= E^{kin}_{A,1}  + E^{r}_a  
 -  \hbar  \,  \vec{v}_{A,1} . \vec{k}_{a} =  E^{kin}_{A,1}  + E^{r}_a  - 2 \, \cos \theta_{A,1} \sqrt{E^{kin}_{A,1}} \, \sqrt{E^{r}_a}
\end{eqnarray}
with the recoil energy defined as
\begin{eqnarray}\label{recoil}
E^{r}_a = \frac{\hbar^2 k_{a}^2}{2m} = \frac{\hbar^2 \omega_{a}^2}{2m c^2} = 
\hbar \omega_{a} \times \frac{\hbar \omega_{a}}{2m c^2}. 
\end{eqnarray}
We used the fact that $\vec{v}_{A,1}  = \frac{\hbar \vec{k}_{A,1}}{m}$
and $\frac{\hbar^2 \vec{k}_{A,1}^2}{2m} = E^{kin}_{A,1}$.
The kinetic energy of the molecule $A$ in state $\ket{0}$
is the sum of the 
kinetic energy of the molecule $A$ in state $\ket{1}$,
the recoil energy $E^r_a$ and the Doppler effect term 
$ -  \hbar  \,  \vec{v}_{A,1} . \vec{k}_{a}$
(see for example \cite{Cohen-Tannoudji_Book_3_2019}).
Using Eq.~\eqref{energy01}, we get that
the energy of the emitted photon by molecule $A$
in the space-fixed frame  (noted $\hbar \omega_{A}^{sf}$) should satisfy
\begin{eqnarray}\label{labframe01}
\hbar \omega_{A}^{sf} =  \varepsilon_1 - \varepsilon_0 
- E^{r}_{a} + \hbar  \,  
\vec{v}_{A,1}.\vec{k}_{a}.
\end{eqnarray}
Let's also consider a general case with a molecule $B$
in state $\ket{1}$ with wavevector $\vec{k}_{B,1}$
undergoing a stimulated absorption process $\ket{1}\rightarrow \ket{2}$ 
in state $\ket{2}$ getting a wavevector $\vec{k}_{B,2}$,
with a photon $b$ of energy 
$\hbar \omega_{b}$ and momentum $ \vec{k}_{b}$.
Similarly, we have
\begin{equation}\label{energy12}
\frac{\hbar^2k^2_{B,2}}{2m}+\varepsilon_{2}=\frac{\hbar^2k^2_{B,1}}{2m}+\varepsilon_{1}+\hbar\omega_B
\end{equation}
and
\begin{eqnarray}\label{momentum12}
\hbar \vec{k}_{B,2} = \hbar \vec{k}_{B,1} + \hbar \vec{k}_{b} .
\end{eqnarray}
We deduce
\begin{eqnarray}\label{dev2}
\frac{\hbar^2 \vec{k}_{B,2}^2}{2m}  &=& \frac{\hbar^2 \big(\vec{k}_{B,1} + \vec{k}_{b}\big)^2 }{2m} 
= E^{kin}_{B,1}  + E^{r}_b  
 +  \hbar  \,  \vec{v}_{B,1} . \vec{k}_{b} =
 E^{kin}_{B,1}  + E^{r}_b  + 2 \, \cos \theta_{B,1} \sqrt{E^{kin}_{B,1}} \, \sqrt{E^{r}_b}
\end{eqnarray}
and
\begin{eqnarray}\label{labframe12}
\hbar \omega_{B}^{sf} =  \varepsilon_2 - \varepsilon_1  + E^{r}_{b} + \hbar  \,  
\vec{v}_{B,1}.\vec{k}_{b}
\end{eqnarray}
the energy of the photon absorbed by the molecule $B$
in the space-fixed frame.
In this study at the F{\"o}rster resonnance, we have 
$\varepsilon_2 - \varepsilon_1 = \varepsilon_1 - \varepsilon_0 \simeq k_B \times 200.6$~mK 
for $^{40}$K$^{87}$Rb molecules. 
As the microwave processes are resonant 
($\Delta=0$) and as the transitions
$\ket{1}\rightarrow \ket{0}$ 
and $\ket{1}\rightarrow \ket{2}$ 
have the same energies, we have 
$\hbar \omega_{a} = \hbar \omega_{b} \simeq k_B \times 200.6$~mK ($\simeq h \times 4.2$~GHz).
We can deduce that the ratio $\frac{\hbar \omega_{a,b}}{2m c^2} \simeq 7 \ 10^{-17}$ in Eq.~\eqref{recoil},
so that $E^{r}_{a,b} \simeq 10^{-14}$~K.
We can also deduce the effect of the Doppler effect
characterized by the terms $2 \, \sqrt{E^{kin}_{A,1}} \, \sqrt{E^{r}_{a}} = 2 \, \sqrt{E^{kin}_{B,1}} \, \sqrt{E^{r}_{b}} \simeq k_B \times 10^{-10}$~K in 
Eq.~\eqref{dev1} or Eq.~\eqref{dev2},
choosing $E^{kin}_{A,1} = E^{kin}_{B,1} \simeq k_B \times 500$~nK and taking $|\cos \theta_{A,1}| = |\cos \theta_{B,1}| =1$. 
We can see that we can neglect both recoil energy and Doppler effect in this study
when compared to the kinetic energies.
In other words, the momentum of a microwave photon is small enough that it does not impact the one of the molecule when an emission or absorption takes place. This implies that the wavevector and kinetic energy
of a molecule excited in state
$\ket{2}$ or de-excited in state $\ket{0}$ are the same
as a molecule in state $\ket{1}$, so that 
$\vec{k}_2 \simeq \vec{k}_0 \simeq \vec{k}_1$ and 
$\hbar^2k^2_2/2m \simeq \hbar^2k^2_0/2m \simeq \hbar^2k^2_1/2m$.

\subsection{Wavevectors of the center-of-mass and relative motions for the combined molecular states}

Let's consider now a given colliding pair with a molecule $A$ in state $i=0,1,2$ 
and a molecule $B$ in state $j=0,1,2$. 
The wavevectors for their center-of-mass and relative motion are given by
\begin{align}\label{wvij}
\vec{K}_{ij} &= \vec{k}_{A,i} + \vec{k}_{B,j}  & \vec{k}_{ij} & =  \big( \vec{k}_{A,i} - \vec{k}_{B,j} \big) / 2
\end{align}
starting from their individual wavevectors defined in the previous section.
For $ij=11$, we have
\begin{align}\label{wv11}
\vec{K}_{11} &= \vec{k}_{A,1} + \vec{k}_{B,1}  & \vec{k}_{11} & =  \big( \vec{k}_{A,1} - \vec{k}_{B,1} \big) / 2
\end{align}
and for $ij=02$, we have
\begin{align}\label{wv02}
\vec{K}_{02} &= \vec{k}_{A,0} + \vec{k}_{B,2} = \vec{k}_{A,1} - \vec{k}_{a} + \vec{k}_{B,1} + \vec{k}_{b} = \vec{K}_{11} + \vec{k}_{b} - \vec{k}_{a}  & \vec{k}_{02}  & =  \big( \vec{k}_{A,0} - \vec{k}_{B,2} \big) /2  = \vec{k}_{11} - \big( \vec{k}_a + \vec{k}_b \big) /2
\end{align}
using Eq.~\eqref{momentum01} and Eq.~\eqref{momentum12}.
Similarly, we have
\begin{align}\label{wvother}
\vec{K}_{00} &= \vec{k}_{A,0} + \vec{k}_{B,0} = \vec{k}_{A,1} - \vec{k}_{a} + \vec{k}_{B,1} - \vec{k}_{a} = \vec{K}_{11} - 2 \vec{k}_{a}  & \vec{k}_{00}  & =  \big( \vec{k}_{A,0} - \vec{k}_{B,0} \big) /2  = \vec{k}_{11}  \nonumber \\
\vec{K}_{01} &= \vec{k}_{A,0} + \vec{k}_{B,1} = \vec{k}_{A,1} - \vec{k}_{a} + \vec{k}_{B,1} = \vec{K}_{11} - \vec{k}_{a}  & \vec{k}_{01}  & = \big( \vec{k}_{A,0} - \vec{k}_{B,1} \big) /2  = \vec{k}_{11} - \vec{k}_{a}/2  \nonumber \\
\vec{K}_{12} &= \vec{k}_{A,1} + \vec{k}_{B,2} = \vec{k}_{A,1} + \vec{k}_{B,1} + \vec{k}_{b} = \vec{K}_{11} + \vec{k}_{b}  & \vec{k}_{12}  & = \big( \vec{k}_{A,1} - \vec{k}_{B,2} \big) /2  = \vec{k}_{11} - \vec{k}_{b}/2  \nonumber \\
\vec{K}_{22} &= \vec{k}_{A,2} + \vec{k}_{B,2} = \vec{k}_{A,1} + \vec{k}_{b} + \vec{k}_{B,1} + \vec{k}_{b} = \vec{K}_{11} + 2 \vec{k}_{b}  & \vec{k}_{22}  & =  \big( \vec{k}_{A,2} - \vec{k}_{B,2} \big) /2  = \vec{k}_{11}  .
\end{align}
We can see that in a general case, if $\vec{k}_a = \vec{k}_b$, that is if the microwave source for the photons $a$ and $b$ is the same (same energy and same wavevector), becoming a unique source of photon, then $\vec{K}_{02} = \vec{K}_{11}$. The center-of-mass motions are then equal for these two types of pairs.
If the recoil energy and the Doppler term can be neglected, 
it is also straightforward that all $\vec{K}_{ij}$ (noted $\vec{K}$ then) and $\vec{k}_{ij}$ (noted $\vec{k}$ then) are mainly identical.
All pairs after the microwave preparation have mainly the same center-of-mass and collide with the same collision energy and wavevector as the initial colliding pair in state $\ket{11}$.

\subsection{Light-matter Hamiltonian. Preparation coefficients for an individual molecule}

The total Hamiltonian describing the light-matter Hamiltonian is given by
$H(t) = H_0 + H'(t)$.
$H_0$ is the time-independent Hamiltonian of the unperturbated system,
including its kinetic energy and its internal energy, 
with eigenfunctions
$\ket{\bar{i}} \equiv \ket{i} e^{\text{i} \vec{k}_i.\vec{\rho}}$ associated to the eigenvalues 
$\hbar\omega_i=\hbar^2k^2_i/2m+\varepsilon_i$ at the F{\"o}rster resonance.
The three states taken into consideration are $i=0,1,2$. Higher states $i>2$ are higher in energy 
and never come into play in the process.
The wavefunctions associated to each of these stationary states are given by 
$\ket{\Psi_i} = \ket{\bar{i}} e^{- \text{i} \omega_i t}$.
$H'(t)$ is the time-dependent Hamiltonian representing the usual light-matter interaction.
Following the previous section, we will assume a unique photon of frequency $\omega_p = \omega_a = \omega_b$ 
and intensity defined by an ac electric field $E_{ac}$.
We consider linear polarization. The Hamiltonian is given by
$ H'(t)=-\vec{d}.\vec{E}_{ac}\cos(\omega_p t)=-\vec{d}.\vec{E}_{ac} \, (e^{+\text{i}\omega_p t}+e^{-\text{i}\omega_p t}) / 2 $.
The total Hamiltonian is given in the basis of the stationary states by 
\cite{Gaubatz_JCP_92_5363_1990,Bergmann_RMP_70_1003_1998,Vitanov_RMP_89_015006_2017}
\begin{equation}
\begin{aligned}
\mathbf{H}(t) & =
\begin{pmatrix}
\hbar\omega_0 & -\frac{\hbar\Omega_{10}}{2} e^{i\omega_p t} & 0 \\

-\frac{\hbar\Omega_{10}}{2}e^{-i\omega_p t} & \hbar\omega_1 & -\frac{\hbar\Omega_{12}}{2}e^{i\omega_p t} \\

0 & -\frac{\hbar\Omega_{12}}{2}e^{-i\omega_p t} & \hbar\omega_2
\end{pmatrix}
\end{aligned}
\end{equation}
where we defined the Rabi frequencies 
$\Omega_{10} = \tilde{d}^{1\leftrightarrow 0} \, E_{ac} / \hbar \equiv \Omega$, 
$\Omega_{12} = \tilde{d}^{1\leftrightarrow 2} \, E_{ac} / \hbar \equiv \chi \, \Omega$,
and the generalized induced dipole moments
$\tilde{d}^{1\leftrightarrow 0} = \bra{0} \vec{d} \ket{1} = \bra{1} \vec{d} \ket{0}$,
$\tilde{d}^{1\leftrightarrow 2} = \bra{2} \vec{d} \ket{1} = \bra{1} \vec{d} \ket{2}$,
computed for example in \cite{Lassabliere_PRA_106_033311_2022} at the F{\"o}rster resonance.
We also note $\Omega_{10} \equiv \Omega$ and $\Omega_{12} \equiv \chi \, \Omega$.
Using now the usual dressed picture of a molecule with a photon for a ladder scheme, we get \cite{Vitanov_RMP_89_015006_2017}
\begin{equation}\label{Hamdressed}
\begin{aligned}
\mathbf{H}(t) & =
\begin{pmatrix}
\hbar\omega_0 + \hbar\omega_p & -\frac{\hbar\Omega_{10}}{2}  & 0 \\

-\frac{\hbar\Omega_{10}}{2} & \hbar\omega_1  & -\frac{\hbar\Omega_{12}}{2} \\

0 & -\frac{\hbar\Omega_{12}}{2} & \hbar\omega_2 - \hbar\omega_p
\end{pmatrix}
=
\begin{pmatrix}
0 & -\frac{\hbar\Omega_{10}}{2}  & 0 \\

-\frac{\hbar\Omega_{10}}{2} & - \hbar \Delta & -\frac{\hbar\Omega_{12}}{2} \\

0 & -\frac{\hbar\Omega_{12}}{2} & - 2 \hbar \Delta
\end{pmatrix} 
\equiv
\begin{pmatrix}
0 & -\frac{\hbar\Omega}{2}  & 0 \\

-\frac{\hbar\Omega}{2} & - \hbar \Delta & - \chi \, \frac{  \hbar\Omega}{2} \\

0 & - \chi \, \frac{\hbar  \Omega}{2} & - 2 \hbar \Delta
\end{pmatrix} .
\end{aligned}
\end{equation}
We defined the detuning $\hbar \Delta = \hbar\omega_p - (\hbar\omega_1 - \hbar\omega_0)$ and 
used the fact that at the F{\"o}rster resonance, we have $\hbar\omega_2 - \hbar\omega_1$ 
= $\hbar\omega_1 - \hbar\omega_0$ since we have 
$\varepsilon_2 - \varepsilon_1$ = $\varepsilon_1 - \varepsilon_0$ for the internal energies,
and $\hbar^2k^2_2/2m \simeq \hbar^2k^2_0/2m \simeq \hbar^2k^2_1/2m$ for the kinetic energies. 
Because of $H'(t)$, the general wavefunction becomes a linear combination of the stationnary states.
It can also be recast in term of new eigenstates $\ket{a_{0,\pm}}$,
the ones that diagonalize Eq.~\eqref{Hamdressed} when $\Delta =0$, the case studied in the paper.
We have then
\begin{eqnarray}\label{qs1MOLSM}
 \ket{\Psi_{mol}}= \sum_{i=0}^2
 {c}_i \ket{\bar{i}}   e^{- \text{i} \omega_i t} 
  = \sum_{j=0,\pm}
 {c}_{a_j}  \ket{a_j}  e^{ -\text{i}  \omega^{a_j} t} 
\end{eqnarray}
with \cite{Gaubatz_JCP_92_5363_1990}
\begin{align}\label{eigena}
\ket{a_0} & = \cos\Theta \ket{\bar{0}}-\sin\Theta\ket{\bar{2}} &
\ket{a_\pm} & = \frac{1}{\sqrt{2}}\bigg(\sin\Theta \ket{\bar{0}}\pm\ket{\bar{1}}+\cos\Theta \ket{\bar{2}}\bigg).
\end{align}
The associated eigenvalues are
\begin{align}
	E^{a_0} & = \hbar\omega^{a_0}=0 & E^{a_\pm} & = \hbar\omega^{a_\pm} = \pm \frac{\hbar}{2}\Omega_{RMS} &
	\qquad \text{with} \qquad 
	\Omega_{RMS} &= \sqrt{\Omega_{10}^2+\Omega_{12}^2} = \Omega \sqrt{1+\chi^2}.
\end{align}
The angle $\Theta$ is defined as $\tan\Theta = \Omega_{10} / \Omega_{12} = 1 / \chi$.
By solving the time-dependent Schr{\"o}dinger equation using the basis of eigenstates 
Eq.~\eqref{eigena} and the third term of Eq.~\eqref{qs1MOLSM}, 
we found that $d{c}_{a_j}/dt=0$ so that ${c}_{a_j}$ are constants and independent of time.
Using the initial conditions $c_1=1$, $c_0=0$, $c_2=0$ in Eq.~\eqref{qs1MOLSM} at $t=0$,
and noting that $\ket{\bar{1}} = \{ \ket{a_+} - \ket{a_-} \} / \sqrt{2}$ in Eq.~\eqref{eigena}, 
we found ${c}_{a_0}= 0$, ${c}_{a_\pm} = \pm 1/{\sqrt{2}}$.
From Eq.~\eqref{eigena}, we have the general relation
\begin{equation}
	\begin{pmatrix}
	{c}_{a_+} \, e^{-\text{i} \omega^{a_+}  t} \\ {c}_{a_0} \\ {c}_{a_-} \, e^{-\text{i} \omega^{a_-}  t}
	\end{pmatrix}
	=
	M
	\begin{pmatrix}
	{c}_{0} \, e^{-\text{i} \omega_0  t} \\ {c}_{1} \, \, e^{-\text{i} \omega_1  t} \\ {c}_{2} \, e^{-\text{i} \omega_2  t}
	\end{pmatrix}
	\equiv
	\begin{pmatrix}
	\frac{\sin\Theta}{\sqrt{2}} & \frac{1}{\sqrt{2}} & \frac{\cos\Theta}{\sqrt{2}} \\
	\cos\Theta & 0 & -\sin\Theta \\
	\frac{\sin\Theta}{\sqrt{2}} & - \frac{1}{\sqrt{2}} & \frac{\cos\Theta}{\sqrt{2}}
	\end{pmatrix}
	\begin{pmatrix}
	{c}_{0} \, e^{-\text{i} \omega_0  t} \\ {c}_{1} \, \, e^{-\text{i} \omega_1  t} \\ {c}_{2} \, e^{-\text{i} \omega_2  t}
	\end{pmatrix}
 \label{equ_M_ini}
\end{equation}
and the inverse one
\begin{equation}
	\begin{pmatrix}
	{c}_{0} \, e^{-\text{i} \omega_0  t} \\ {c}_{1} \, \, e^{-\text{i} \omega_1  t} \\ {c}_{2} \, e^{-\text{i} \omega_2  t}
	\end{pmatrix}
	=
	{M}^{-1}
	\begin{pmatrix}
	{c}_{a_+} \, e^{-\text{i} \omega^{a_+}  t} \\ {c}_{a_0} \\ {c}_{a_-} \, e^{-\text{i} \omega^{a_-}  t}
	\end{pmatrix}
	\equiv
	\begin{pmatrix}
	\frac{\sin\Theta}{\sqrt{2}} &  \cos\Theta & \frac{\sin\Theta}{\sqrt{2}} \\
	\frac{1}{\sqrt{2}}& 0 &  - \frac{1}{\sqrt{2}}  \\
	\frac{\cos\Theta}{\sqrt{2}}  & -\sin\Theta & \frac{\cos\Theta}{\sqrt{2}}
	\end{pmatrix}
	\begin{pmatrix}
	{c}_{a_+} \, e^{-\text{i} \omega^{a_+}  t} \\ {c}_{a_0} \\ {c}_{a_-} \, e^{-\text{i} \omega^{a_-}  t}
	\end{pmatrix} \\
		=
	\begin{pmatrix}
	\frac{\sin\Theta}{\sqrt{2}} &  \cos\Theta & \frac{\sin\Theta}{\sqrt{2}} \\
	\frac{1}{\sqrt{2}}& 0 &  - \frac{1}{\sqrt{2}}  \\
	\frac{\cos\Theta}{\sqrt{2}}  & -\sin\Theta & \frac{\cos\Theta}{\sqrt{2}}
	\end{pmatrix}
	\begin{pmatrix}
	\frac{1}{\sqrt2} \, e^{-\text{i} \omega^{a_+}  t} \\ 0 \\ -\frac{1}{\sqrt2}  \, e^{-\text{i} \omega^{a_-}  t}
	\end{pmatrix} .
 \label{equ_M_inv}
\end{equation}
Then
\begin{equation}
	\begin{pmatrix}
	{c}_{0} \, e^{-\text{i} \omega_0  t} \\ {c}_{1} \, \, e^{-\text{i} \omega_1  t} \\ {c}_{2} \, e^{-\text{i} \omega_2  t}
	\end{pmatrix}
	=
	\begin{pmatrix}
	- \text{i} \sin\Theta \, \sin\bigg(\sqrt{1+\chi^2} \, \Omega t /2 \bigg) \\  \cos\bigg(\sqrt{1+\chi^2} \, \Omega  t/2 \bigg)  \\ - \text{i} \cos\Theta \, \sin\bigg(\sqrt{1+\chi^2} \, \Omega t /2 \bigg) 
	\end{pmatrix} .
 \label{equ_c012}
\end{equation}
At time $\tau$, we have then
\begin{align}
{c}_{0} &=  - \text{i} \sin\Theta \, \sin\eta \, e^{\text{i} \omega_0  \tau} & 
{c}_{1} &=  \cos\eta \, e^{\text{i} \omega_1  \tau} &
{c}_{2} &=  - \text{i} \cos\Theta \, \sin\eta \, e^{\text{i} \omega_2  \tau}
 \label{equ_c012_attau}
\end{align}
using the notation $\eta = \sqrt{1+\chi^2} \, \Omega \tau /2$.
If we set ${c}_i = C_i \, e^{\text{i} \omega_i \tau}$, we identify
\begin{align}  \label{equ_BigC012_attau}
{C}_{0} &=  - \text{i} \sin\Theta \sin\eta & 
{C}_{1} &=  \cos\eta &
{C}_{2} &=  - \text{i} \cos\Theta \sin\eta.
\end{align}

\subsection{Preparation coefficients for the combined molecular states}

Using Eq. 1 of the paper, we get for the overall incident collisional wavefunction between two molecules
\begin{eqnarray}\label{CMSnonsym}
 \ket{\Psi_{col}^{inc}} &=& \ket{\Psi_{mol}} \otimes \ket{\Psi_{mol}} \nonumber \\
 &=& C_0^2 e^{i\left( (\vec{k}_0+\vec{k}_0).\vec{\rho}-(\omega_0+\omega_0) (t-\tau) \right) }\ket{00} 
 + C_1^2 e^{i\left( (\vec{k}_1+\vec{k}_1).\vec{\rho}-(\omega_1+\omega_1) (t-\tau) \right) }\ket{11} 
 \nonumber \\
 &+& C_2^2 e^{i\left( (\vec{k}_2+\vec{k}_2).\vec{\rho}-(\omega_2+\omega_2) (t-\tau) \right) }\ket{22} 
  + C_0 C_1 e^{i\left( (\vec{k}_0+\vec{k}_1).\vec{\rho}-(\omega_0+\omega_1) (t-\tau) \right) }\left(\ket{01}+\ket{10}\right) \nonumber \\
  &+& C_1 C_2 e^{i\left( (\vec{k}_1+\vec{k}_2).\vec{\rho}-(\omega_1+\omega_2) (t-\tau) \right) }\left(\ket{12}+\ket{21}\right) 
 + C_0 C_2 e^{i\left( (\vec{k}_0+\vec{k}_2).\vec{\rho}-(\omega_0+\omega_2) (t-\tau) \right) }\left(\ket{02}+\ket{20}\right).
\end{eqnarray}
Since these states need to be properly symmetrized under exchange of identical particles and since 
we start with identical molecules in indistinguishable states, 
only symmetric states are involved in the dynamics 
\cite{Wang_NJP_17_035015_2015}. 
These symmetric states are defined as 
$\ket{ij+}=\frac{1}{\sqrt{2(1+\delta_{i,j})}}\left(\ket{ij}+\ket{ji}\right)$.
Then Eq.~\eqref{CMSnonsym} is rewritten
\begin{eqnarray}\label{CMSsym}
 \ket{\Psi_{col}^{inc}}  &=& C_0^2 e^{i\left( (\vec{k}_0+\vec{k}_0).\vec{\rho}-(\omega_0+\omega_0) (t-\tau) \right) }\ket{00+} 
 + C_1^2 e^{i\left( (\vec{k}_1+\vec{k}_1).\vec{\rho}-(\omega_1+\omega_1) (t-\tau) \right) }\ket{11+} 
 \nonumber \\
 &+& C_2^2 e^{i\left( (\vec{k}_2+\vec{k}_2).\vec{\rho}-(\omega_2+\omega_2) (t-\tau) \right) }\ket{22+} 
  + \sqrt{2} \, C_0 C_1 \, e^{i\left( (\vec{k}_0+\vec{k}_1).\vec{\rho}-(\omega_0+\omega_1) (t-\tau) \right) }\ket{01+} \nonumber \\
  &+& \sqrt{2} \,  C_1 C_2 \, e^{i\left( (\vec{k}_1+\vec{k}_2).\vec{\rho}-(\omega_1+\omega_2) (t-\tau) \right) }\ket{12+} 
 + \sqrt{2} \, C_0 C_2 \, e^{i\left( (\vec{k}_0+\vec{k}_2).\vec{\rho}-(\omega_0+\omega_2) (t-\tau) \right) }\ket{02+}.
\end{eqnarray}
In the paper, we omit the plus sign for the states $\ket{00+}$, $\ket{11+}$ $\ket{22+}$, 
as they are obviously symmetric under exchange of particles.
Then 
\begin{align}  \label{equ_BigCalpha_attau}
C_{00} &= C_0^2 & 
C_{11} &= C_1^2 &
C_{22} &= C_2^2  \nonumber \\
C_{01} &= \sqrt{2} \, C_0 C_1 & C_{12} &= \sqrt{2} \, C_1 C_2 & C_{02} &= \sqrt{2} \, C_0 C_2 .
\end{align}

\subsection{Overall loss rate coefficient and probabilities}

From Eq. 6 of the paper, we can make the link between the $T$ matrix and the $S$ matrix for the interference term
\begin{eqnarray}
T^{qs}_{\alpha' l' m_l',11/02+ \, lm_l} &=& C_{11} T_{\alpha' l' m_l',11 \, lml} + C_{02+} T_{\alpha' l' m_l',02+ \, lm_l} \nonumber \\
&=& C_{11} (S_{\alpha' l' m_l',11 \, lml} - \delta_{\alpha' l' m_l',11 \, lml}) + C_{02+} (S_{\alpha' l' m_l',02+ \, lml} - \delta_{\alpha' l' m_l',02+ \, lml}) \nonumber \\
&=& (C_{11} S_{\alpha' l' m_l',11 \, lml} + C_{02+} S_{\alpha' l' m_l',02+ \, lml}) - (C_{11} \delta_{\alpha' l' m_l',11 \, lml} + C_{02+} \delta_{\alpha' l' m_l',02+ \, lml}) .
\end{eqnarray}
This defines a new expression for the scattering matrix for the interference term
\begin{eqnarray}\label{Sqs}
S^{qs}_{\alpha' l' m_l',11/02+ \, lm_l} &=& C_{11} S_{\alpha' l' m_l',11 \, lml} + C_{02+} S_{\alpha' l' m_l',02+ \, lm_l}.
\end{eqnarray}
For the satellite terms we have simply
\begin{eqnarray}
S^{qs}_{\alpha' l' m_l',\alpha \, lm_l} &= C_{\alpha} S_{\alpha' l' m_l',\alpha \, lml}.
\end{eqnarray}
The modulus square of the $S^{qs}$ matrix elements are associated to the probabilities,
the sum of which for a given column should give unity.
Let's check the conservation of these
associated probabilities.
For a given column of the $S^{qs}$ matrix for a given $l, m_l$,
and for all the states $\alpha'$ (spending all reactants or products),
we have for the interference term
\begin{eqnarray}
\sum_{\alpha', l', m_l'} |S^{qs}_{\alpha' \, l' \, m_l', 11/02+ \, l \, m_l}|^2
 &=&  \sum_{\alpha', l', m_l'} 
 |C_{11} \,  S_{\alpha' \, l' \, m_l',11 \, l \, m_l}
 + C_{02+} \,  S_{\alpha' \, l' \, m_l',02+ \, l \, m_l}|^2 \nonumber \\
 &=& \sum_{\alpha', l', m_l'} 
 \bigg( |C_{11}|^2 \, | S_{\alpha' \, l' \, m_l',11 \, l \, m_l}|^2
 + |C_{02+}|^2 \, | S_{\alpha' \, l' \, m_l',02+ \, l \, m_l}|^2 \nonumber \\
 &+& C_{11}^* \, C_{02+} \, [S_{\alpha' \, l' \, m_l',11 \, l \, m_l}]^* \,  S_{\alpha' \, l' \, m_l',02+ \, l \, m_l} + C_{02+}^* \, C_{11} \, [S_{\alpha' \, l' \, m_l',02+ \, l \, m_l}]^* \,  S_{\alpha' \, l' \, m_l',11 \, l \, m_l} \bigg) \nonumber \\
  &=&  |C_{11}|^2 \, \sum_{\alpha', l', m_l'} \, | S_{\alpha' \, l' \, m_l',11 \, l \, m_l}|^2
 + |C_{02+}|^2 \, \sum_{\alpha', l', m_l'} \, | S_{\alpha' \, l' \, m_l',02+ \, l \, m_l}|^2 \nonumber \\
   &=&  |C_{11}|^2 + |C_{02+}|^2 \nonumber \\
   &\equiv&  P^{el,qs}_{11/02+ \, l \, m_l} + P^{in,qs}_{11/02+ \, l \, m_l} + P^{re,qs}_{11/02+ \, l \, m_l} .
\end{eqnarray}
This is equal to the sum of the elastic $P^{el,qs}$, inelastic $P^{in,qs}$ and reactive 
$P^{re,qs}$ probabilities starting in the interfering state $11/02+$.
We used the fact that a $S$ matrix is unitary so that we have
\begin{eqnarray}
 \sum_{\alpha', l', m_l'} [S_{\alpha' \, l' \, m_l',11 \, l \, m_l}]^* \,  S_{\alpha' \, l' \, m_l',02+ \, l \, m_l} &=&  \sum_{\alpha', l', m_l'} [S_{\alpha' \, l' \, m_l', 02+ \, l \, m_l}]^* \,  S_{\alpha' \, l' \, m_l', 11 \, l \, m_l} = 0 \nonumber \\
\sum_{\alpha', l', m_l'} 
 | S_{\alpha' \, l' \, m_l',11 \, l \, m_l}|^2
 &=& \sum_{\alpha', l', m_l'} | S_{\alpha' \, l' \, m_l',02+ \, l \, m_l}|^2 = 1 . 
\end{eqnarray}
Similarly for the satellite terms, we have
\begin{eqnarray}
\sum_{\alpha', l', m_l'} |S^{qs}_{\alpha' \, l' \, m_l', \alpha \, l \, m_l}|^2
 &=&   |C_{\alpha}|^2  \equiv  P^{el,qs}_{\alpha \, l \, m_l} + P^{in,qs}_{\alpha \, l \, m_l} + P^{re,qs}_{\alpha \, l \, m_l} .
\end{eqnarray}
This is equal to the sum of the elastic $P^{el,qs}$, inelastic $P^{in,qs}$ and reactive 
$P^{re,qs}$ probabilities starting in the satellite states $\alpha=00, 01+, 12+, 22$.
If we sum all of those interference and satellite probabilities for a given $l, m_l$
over $\alpha = 11/02+, 00, 01+, 12+, 22$, we have then that
\begin{eqnarray}
\sum_{\alpha}^* \, \sum_{\alpha', l', m_l'} |S^{qs}_{\alpha' \, l' \, m_l', \alpha \, l \, m_l}|^2
 &=&  \sum_{\alpha}^* \, |C_{\alpha}|^2 =  |C_0|^4 + |C_1|^4 + |C_2|^4 + 2 |C_0|^2 |C_1|^2 + 2 |C_1|^2 |C_2|^2 + 2 |C_0|^2 |C_2|^2 \nonumber \\
 &=& (|C_0|^2 + |C_1|^2 + |C_2|^2)^2 = 1 
\end{eqnarray}
where we used Eq.~\eqref{equ_BigCalpha_attau} and  Eq.~\eqref{equ_BigC012_attau}.
We then checked that the new expression of the $S$ matrices in Eq.~\eqref{Sqs}
provides a sum of all probabilities of unity (for a given $l, m_l$), as it should be.
\\

Among all the states $\alpha'$, we don't have information on those which correspond to the products of the reaction. Instead, the computed $S$ matrix is sub-unitary (due to an appropriate short-range absorbing boundary condition as explained in \cite{Wang_NJP_17_035015_2015})
so that the difference with  unitarity provides the overall phenomenological reactive probability. Then for a given $\alpha \,l \, m_l$, we have instead for the interference term
\begin{eqnarray}
\sum^{el+in}_{\alpha', l', m_l'} |S^{qs}_{\alpha' \, l' \, m_l', 11/02+ \, l \, m_l}|^2
 &\equiv&  P^{el,qs}_{11/02+ \, l \, m_l} + P^{in,qs}_{11/02+ \, l \, m_l} \nonumber \\
 & = & \sum^{el+in}_{\alpha', l', m_l'} 
 |C_{11} \,  S_{\alpha' \, l' \, m_l',11 \, l \, m_l}
 + C_{02+} \,  S_{\alpha' \, l' \, m_l',02+ \, l \, m_l}|^2 \nonumber \\
 &=& \sum^{el+in}_{\alpha', l', m_l'} 
 \bigg( |C_{11}|^2 \, | S_{\alpha' \, l' \, m_l',11 \, l \, m_l}|^2
 + |C_{02+}|^2 \, | S_{\alpha' \, l' \, m_l',02+ \, l \, m_l}|^2 \nonumber \\
 &+& C_{11}^* \, C_{02+} \, [S_{\alpha' \, l' \, m_l',11 \, l \, m_l}]^* \,  S_{\alpha' \, l' \, m_l',02+ \, l \, m_l} + C_{02+}^* \, C_{11} \, [S_{\alpha' \, l' \, m_l',02+ \, l \, m_l}]^* \,  S_{\alpha' \, l' \, m_l',11 \, l \, m_l} \bigg) \nonumber \\
  &=&  |C_{11}|^2 \, \sum^{el+in}_{\alpha', l', m_l'} \, | S_{\alpha' \, l' \, m_l',11 \, l \, m_l}|^2
 + |C_{02+}|^2 \, \sum^{el+in}_{\alpha', l', m_l'} \, | S_{\alpha' \, l' \, m_l',02+ \, l \, m_l}|^2 \nonumber \\
   &=&  |C_{11}|^2 \,   (P^{el}_{11 \, l \, m_l} + P^{in}_{11 \, l \, m_l}) 
 + |C_{02+}|^2 \,  (P^{el}_{02+ \, l \, m_l} + P^{in}_{02+ \, l \, m_l})  \nonumber \\
    &=&  |C_{11}|^2 \,   (1 - P^{re}_{11 \, l \, m_l}) 
 + |C_{02+}|^2 \,  (1 - P^{re}_{02+ \, l \, m_l} )  \nonumber \\
   &=&  |C_{11}|^2 + |C_{02+}|^2 - |C_{11}|^2 \, P^{re}_{11 \, l \, m_l} - |C_{02+}|^2 \,  P^{re}_{02+ \, l \, m_l}
\end{eqnarray}
where the special sum indicates that the sum on $\alpha'$ is done on the elastic and inelastic states only.
We can then identify
\begin{eqnarray}
P^{re,qs}_{11/02+ \, l \, m_l} = |C_{11}|^2 \, P^{re}_{11 \, l \, m_l} + |C_{02+}|^2 \,  P^{re}_{02+ \, l \, m_l}
\end{eqnarray}
from the fact that
$P^{el,qs}_{11/02+ \, l \, m_l} + P^{in,qs}_{11/02+ \, l \, m_l} + P^{re,qs}_{11/02+ \, l \, m_l} = |C_{11}|^2 + |C_{02+}|^2 $. 
For the satellite terms, we have 
\begin{eqnarray}
\sum^{el+in}_{\alpha', l', m_l'} |S^{qs}_{\alpha' \, l' \, m_l', \alpha \, l \, m_l}|^2 
&\equiv&  P^{el,qs}_{\alpha \, l \, m_l} + P^{in,qs}_{\alpha \, l \, m_l} \nonumber  
=  \sum^{el+in}_{\alpha', l', m_l'} 
 |C_{\alpha} \,  S_{\alpha' \, l' \, m_l', \alpha \, l \, m_l}|^2 
  =  |C_{\alpha}|^2 \,   (P^{el}_{\alpha \, l \, m_l} + P^{in}_{\alpha \, l \, m_l})
   =  |C_{\alpha}|^2 \,   (1 - P^{re}_{\alpha \, l \, m_l})  .
\end{eqnarray}
Then we identify
\begin{eqnarray}
P^{re,qs}_{\alpha \, l \, m_l} = |C_{\alpha}|^2 \, P^{re}_{\alpha \, l \, m_l} 
\end{eqnarray}
from the fact that
$P^{el,qs}_{\alpha \, l \, m_l} + P^{in,qs}_{\alpha \, l \, m_l} + P^{re,qs}_{\alpha \, l \, m_l} = |C_{\alpha}|^2 $. 

\twocolumngrid

\end{document}